\documentclass[english,aps,manuscript]{revtex4}
\usepackage[T1]{fontenc}
\usepackage[latin9]{inputenc}
\usepackage{graphicx}
\usepackage{color}
\usepackage{amsmath}
\usepackage{epsfig}
\makeatletter
\@ifundefined{textcolor}{}
{%
 \definecolor{BLACK}{gray}{0}
 \definecolor{WHITE}{gray}{1}
 \definecolor{RED}{rgb}{1,0,0}
 \definecolor{GREEN}{rgb}{0,1,0}
 \definecolor{BLUE}{rgb}{0,0,1}
 \definecolor{CYAN}{cmyk}{1,0,0,0}
 \definecolor{MAGENTA}{cmyk}{0,1,0,0}
 \definecolor{YELLOW}{cmyk}{0,0,1,0}
}

\makeatother

\usepackage{babel}
\begin{document}

\title{Dynamics of Polymers: a Mean-Field Theory}
\author{Glenn~H.~Fredrickson}
\affiliation{Department of Chemical Engineering, University of California, Santa Barbara, California 93106, USA}
\affiliation{Materials Research Laboratory, University of California, Santa Barbara, California 93106, USA}
\affiliation{Department of Materials, University of California, Santa Barbara, California 93106, USA}
\author{Henri Orland}
\affiliation{Institut de Physique Th\'eorique, CE-Saclay, CEA, F-91191 Gif-sur-Yvette Cedex, France}
\date{\today}

\begin{abstract}
We derive a general mean-field theory of inhomogeneous polymer dynamics; a theory whose form has been speculated and widely applied, but not heretofore derived. Our approach involves a functional integral representation of a  Martin-Siggia-Rose type description of the exact many-chain dynamics. A saddle point approximation to the generating functional, involving conditions where the MSR action is stationary with respect to a collective density field $\rho$ and a conjugate MSR response field $\phi$, produces the desired dynamical mean-field theory. Besides clarifying the proper structure of mean-field theory out of equilibrium, our results have implications for numerical studies of polymer dynamics involving hybrid particle-field simulation techniques such as the single-chain in mean-field method (SCMF).

\end{abstract}

\maketitle

\section{Introduction}

Our understanding of polymer dynamics has advanced considerably with the advent of intuitive mean-field concepts such as the tube model\cite{pgg,doiedwards}, which have been extensively exploited to derive molecularly-inspired constitutive laws for entangled polymers. These constructs have been remarkably successful at predicting and reproducing a wide range of linear and nonlinear rheological phenomena in homogeneous polymeric fluids. Nonetheless, the dynamic properties of the mean-field, e.g. the tube, are postulated, rather than derived from first principles, so there remain some unsatisfying aspects to the theory.

The situation in \emph{inhomogeneous} polymeric liquids is far worse\cite{ghf02}; the most sophisticated mean-field constructs, such as the two-fluid model of Doi and Onuki\cite{doionuki}, allow for coupled equations for collective densities and stresses to be derived, but give little guidance into the constitutive laws relating stress and flow, and especially across steep gradients in composition. Even simpler mean-field theories such as the dynamic density functional theory (DDFT) method of Fraaije and co-workers\cite{fraaije93,fraaije97} do not even engage stress (or recoverable strain) and momentum density as collective dynamic variables, but retain only monomer densities evolved by Fickian dynamics driven by chemical potential gradients computed with static self-consistent field theory (SCFT). Similar approaches have been proposed by Hasegawa and Doi\cite{hasegawa}, Yeung and Shi\cite{yeung}, Reister et. al.\cite{reister}, and M\"{u}ller and Schmid\cite{schmid}, among others. In such a highly simplified framework, one can at best hope for a qualitative description of the low-frequency, long-wavelength, quiescent response of the fluid, embedding molecular details such as degree of entanglement, molecular weight and architecture, and monomeric friction into one or more phenomenological Onsager kinetic coefficients.

In principle, Mori-Zwanzig type projection operator methods\cite{zwanzig} can be used to project out linear or nonlinear dynamical equations for a set of collective fields (i.e. densities, stresses, etc.) from a microscopic Newtonian or Brownian many-chain dynamics. However, the reduction in dimension associated with the projection process introduces memory kernels in the collective equations that are essentially intractable, i.e. as difficult as the starting microscopic dynamics. Simplifying approximations to the kernels, such as the local-equilibrium, Markov approximation of Kawasaki and Sekimoto\cite{kawasaki}, are uncontrolled, unpredictable, and rely on the retained fields being the only slow collective modes in the fluid -- an unlikely proposition.

Recently, a new class of polymer simulation techniques has evolved that is a hybrid between particle-based methods and field-based methods. The most widely used variant is the single-chain in mean-field (SCMF) approach pioneered by M\"{u}ller and co-workers\cite{mueller1,mueller2} in which discrete polymer chains are moved independently in dynamical mean-fields, the fields updated periodically by using the instantaneous microscopic densities implied by the chain monomer coordinates and assuming the same local and instantaneous relation between densities and fields that holds in mean-field theory and at equilibrium (i.e. in SCFT). Daoulas and Muller \cite{mueller2} have argued that the SCMF, by virtue of using the instantaneous microscopic densities to construct the fields, actually produces equilibrium results that go beyond SCFT by including field fluctuations. They have supported this argument by showing good qualitative agreement with full Monte Carlo (MC) simulations of the equivalent particle model, although quantitative agreement requires frequent updating of the densities, which reduces the computational advantage relative to full simulations. \emph{Away from equilibrium}, there is also the question of the validity of the local and instantaneous connection between densities and fields. If fields are updated less frequently than the evolving non-interacting chains using SCMF or related techniques\cite{saphia,naray,bonnet,ganesan}, it is not obvious that the equilibrium-inspired field updating procedure is correct (even in a mean-field sense) for systems out of equilibrium.

In the present paper, we show how a dynamical mean-field theory can be rigorously derived for a simple model of flexible homopolymers in an implicit good solvent. The underlying microscopic model is a Rouse-Brownian dynamics for the polymer segments of each chain and for simplicity hydrodynamic interactions and externally imposed flows are not included. Our results show that indeed, at the mean-field level, there is an instantaneous relation between the mean-field and the density that coincides with that employed in the SCMF framework. Furthermore, we observe that in a numerical implementation, chains and fields must be moved simultaneously for exact evolution on the dynamical mean-path.

\section{Microscopic dynamics and functional integral representation}

In the present paper we consider a simple system of $M$ interacting chains, each comprised of $N$ monomers. We assume that the chains are homopolymers
but the generalization to any chemical sequence is straightforward. The chains are embedded in an implicit solvent and we assume that the non-bonded interactions among monomers are pairwise and the pair potential (in units of $k_B T$) is denoted by $v(r-r').$

The Hamiltonian of the system can be written (in the continuous chain representation) as
\[
\beta H=\frac{3}{2a^{2}}\sum_{k=1}^{M}\int_{0}^{N}ds\left(\frac{dr_{k}}{ds}\right)^{2}+\frac{1}{2}\sum_{k,l=1}^{M}\int_{0}^{N}ds\int_{0}^{N}ds'v(r_{k}(s)-r_{l}(s'))
\]
where $a$ is the statistical segment length and $v(r_k(s) - r_l(s'))$ represents the effective interaction of monomer $s$ of chain $k$ with monomer $s'$ of chain $l$.
The corresponding Langevin-Rouse equation (without hydrodynamic flow) reads:
\begin{equation}
\frac{dr_{k}}{dt}=D\beta\left(\frac{3}{a^{2}}\frac{d^{2}r_{k}}{ds^{2}}-\sum_{l}\int_{0}^{N}ds'\nabla_{k}v(r_{k}(s,t)-r_{l}(s',t))\right)+\eta_{k}(s,t)\label{eq:langevin}
\end{equation}
where $D$ is a monomeric diffusion coefficient and the Gaussian white noise satisfies
\begin{eqnarray*}
<\eta_{k}(s,t)> & = & 0    \\
<\eta_{k}(s,t)\eta_{l}(s',t')>&=&2D\delta_{kl}\delta(s-s')\delta(t-t')
\end{eqnarray*}
We note that the form of the interaction potential is arbitrary, but is assumed differentiable. If the potential has a sufficiently hard core, chain crossings would be eliminated, so our microscopic dynamics can capture entanglement effects in spite of the nomenclature ``Rouse.''
{
Introducing the instantaneous monomer density field $\hat\rho(r,t)$
\begin{eqnarray}
\hat\rho(r,t)=\sum_{k=1}^{M}\int_{0}^{N}ds \:\delta(r-r_{k}(s,t))
\end{eqnarray}
the Langevin-Rouse equation can be written, in exact form as
\begin{equation}
\label{density_langevin}
\frac{dr_{k}}{dt}=D\beta\left(\frac{3}{a^{2}}\frac{d^{2}r_{k}}{ds^{2}}-\int dr' \nabla_{k}v(r_{k}(s,t)-r')\hat \rho(r',t) \right)+\eta_{k}(s,t)
\end{equation}
}
We will come back to this equation in the discussion of section \ref{discussion}.
\color{black}

As is well known, there are many ways to discretize a stochastic differential equation such as (\ref{eq:langevin}) in time, for example using the Ito or Stratonovich prescription {(\cite{kleinert})}\color{black}. When expanded consistently in the time step $\varepsilon,$ all such methods
yield an identical continuous time limit theory. In the following, we will adopt the standard Ito discretization (also known as the Euler-Maruyama scheme):
\begin{eqnarray}
\label{langevin}
r_{k}(s,t+\varepsilon)&=&r_{k}(s,t)+D\beta\varepsilon\left(\frac{3}{a^{2}}\frac{d^{2}r_{k}}{ds^{2}}-\sum_{l}\int_{0}^{N}ds'\nabla_{k}v(r_{k}(s,t)-r_{l}(s',t))\right)
\nonumber \\
&+&\sqrt{2D\varepsilon}\zeta_{k}(s,t)
\end{eqnarray}
with the rescaled Gaussian noise $\zeta_{k}$ defined by
\begin{eqnarray*}
<\zeta_{k}(s,t)> & = & 0  \\
<\zeta_{k}(s,t)\zeta_{l}(s',t')>&=&\delta_{kl}\delta(s-s')\delta_{tt'}
\end{eqnarray*}
and where we employ a Kronecker delta function for the discretized time variable $t$.
The distribution function for the
Gaussian noise can be written as
\[
\mathcal{P}(\zeta_{k})=\frac{1}{\mathcal{N}}\exp\left(-\frac{1}{2}\sum_{t}\sum_{k}\int_{0}^{N}ds\zeta_{k}^{2}(s,t)\right)
\]
where $\mathcal{N}$ is a normalization factor.

Using a variant of the Martin-Siggia-Rose (MSR)\cite{msr} formalism due to Jensen\cite{jensen}, a generating functional for the time-discretized dynamics can be written as

\begin{eqnarray}
\label{proba}
 &  & \mathcal{P}(h_{k}(s,t))=\int\mathcal{D}r_{k}(s,t)e^{i\varepsilon\sum_{t}\sum_{k}\int_{0}^{N}dsh_{k}(s,t)r_{k}(s,t)}\nonumber \\
 &  & <\prod_{t}\prod_{k} \prod_{s}\delta\Bigg(r_{k}(s,t+\varepsilon)-r_{k}(s,t)\nonumber \\
 &  & -D\beta\varepsilon\left(\frac{3}{a^{2}}\frac{d^{2}r_{k}}{ds^{2}}-\sum_{l}\int_{0}^{N}ds'\nabla_{k}v(r_{k}(s,t)-r_{l}(s',t))\right)-\sqrt{2D\varepsilon}\zeta_{k}(s,t)\Bigg)>
\end{eqnarray}
where $h_{k}(s,t)$ is a source used to generate expectation values of
$r_{k}(s,t)$ and $<...>$ denotes the average over the Gaussian noise
$\zeta_{k}$. Note that within the present Ito convention, there is no Jacobian (functional determinant) as in the standard MSR method, since the delta-function yields  $r(s,t+\varepsilon)$ explicitly as a function of $r(s,t)$.
This discretization also produces causal propagators in time.

\section{Collective variables and the dynamical mean-field approximation}

We next introduce collective density fields into the framework, similar to an earlier MSR polymer dynamics study by Fredrickson and Helfand\cite{ghf90}. By means of the identity $1 = \int \mathcal{D}\rho \; \delta (\rho - \hat\rho )$, where $\hat\rho(r,t)$ is the microscopic monomer density field
\[
\hat\rho(r,t)=\sum_{k=1}^{M}\int_{0}^{N}ds \:\delta(r-r_{k}(s,t))
\]
followed by an exponential representation of the delta functional, the dynamics can be rewritten as
\begin{eqnarray}
 &  & \mathcal{P}(h_{k}(s,t))=\int\mathcal{D}\rho\mathcal{D}\phi e^{i\varepsilon\sum_{t}\int dr\rho(r,t)\phi(r,t)}\int\mathcal{D}r_{k}(s,t)e^{i\varepsilon\sum_{t}\sum_{k}\int_0^N ds\left(h_{k}(s,t)r_{k}(s,t)-\phi(r_{k}(s,t),t)\right)}\nonumber \\
 &  & <\prod_{t}\prod_{k}\delta\Bigg(r_{k}(s,t+\varepsilon)-r_{k}(s,t)\nonumber \\
 &  & -D\beta\varepsilon\left(\frac{3}{a^{2}}\frac{d^{2}r_{k}}{ds^{2}}-\int dr'\nabla_{k}v(r_{k}(s,t)-r')\rho(r',t)\right)-\sqrt{2D\varepsilon}\zeta_{k}(s,t)\Bigg)>
\end{eqnarray}
We note that the segment density field $\rho$ now appears in the force term involving the pair potential $v$ and that a second collective field $\phi (r ,t)$, arising from the exponentiation of the delta functional, plays the role of an MSR response field as it is conjugate to $\rho$.

In the following, we will assume identical fields $h_{k}=h$ on all
chains. In that case, the $M$ chains are decoupled,
and the generating functional can be written as
\[
\mathcal{P}(h(s,t))=\int\mathcal{D}\rho\mathcal{D}\phi e^{i\varepsilon\sum_{t}\int dr\rho(r,t)\phi(r,t)+M\log Q(\rho,\phi)}
\]
where $Q (\rho , \phi )$ is the MSR generating functional for a \emph{single chain}:
\begin{eqnarray}
\label{Q}
Q(\rho,\phi) & = & \int\mathcal{D}r(s,t)e^{i\varepsilon\sum_{t}\int_0^N ds\left(h(s,t)r(s,t)-\phi(r(s,t),t)\right)}
  <  \prod_{t,s}\delta\Bigg(r(s,t+\varepsilon)-r(s,t) \nonumber \\
 &-&D\beta\varepsilon\left(\frac{3}{a^{2}}\frac{d^{2}r}{ds^{2}}-\int dr'\nabla_{r}v(r(s,t)-r')\rho(r',t)\right)-\sqrt{2D\varepsilon}\zeta(s,t)\Bigg) >
\end{eqnarray}

The above equations represent an exact reformulation of the many-chain dynamics in functional integral form. Previous researchers have arrived at this expression but proceeded differently. Fredrickson and Helfand\cite{ghf90} expanded $Q$ to quadratic order in the fields and showed that this leads to a closed theory for response and space-time correlation functions consistent with the dynamical random phase approximation (RPA)\cite{benoit}. Grzetic\cite{grzetic} reexpressed the single chain Langevin dynamics in Fokker-Planck form, but did not have a strategy for tackling the high dimensional FP equation.

Here we seek a dynamical mean-field approximation by evaluating  $\mathcal{P}$ using the saddle-point (SP) method. For simplicity in the following we will take the source to zero: $h=0.$
The mean-field SP equations result from setting the first variations of the action functional
\[
\mathcal{A} (\rho , \phi ) = i\varepsilon\sum_{t}\int dr\rho(r,t)\phi(r,t)+M\log Q(\rho,\phi)
\]
to zero. The following expressions are obtained:
\begin{equation}
\label{rho}
\rho(r,t)=M\int_{0}^{N}ds<\delta(r(s,t)-r)>_{Q}
\end{equation}
and
\begin{equation}
\label{phi}
\phi(r,t)=i\frac{M}{\varepsilon}\frac{\delta}{\delta\rho(r,t)}\log Q(\rho,\phi)
\end{equation}

The notation $<...>_Q$ denotes an expectation value with respect to the single chain dynamics defined by $Q$, in eq. (\ref{Q}). Using the Fourier representation of the $\delta$-functions, we have
\begin{eqnarray}
Q(\rho,\phi) & = &{ \int\mathcal{D}r(s,t) \mathcal{D}q(s,t)} \color{black} e^{-i\varepsilon\sum_{t}\int ds\phi(r(s,t),t)}\nonumber \\
 & < & \exp\Bigg(i\sum_{t}\int_{0}^{N}dsq(s,t)\Bigg(r(s,t+\varepsilon)-r(s,t)\nonumber \\
 & - & D\beta\varepsilon\left(\frac{3}{a^{2}}\frac{d^{2}r}{ds^{2}}-\int dr'\nabla_{r}v(r(s,t)-r')\rho(r',t)\right)-\sqrt{2D\varepsilon}\zeta(s,t)\Bigg)\Bigg)>
\end{eqnarray}
and thus
\begin{eqnarray}
\frac{\delta Q}{\delta\rho(r,t)} & = & iD\beta\varepsilon\int\mathcal{D}r(s,t) \mathcal{D}q(s,t)e^{-i\varepsilon\sum_{t}\int ds\phi(r(s,t),t)}\int_{0}^{N}dsq(s,t)\nabla v(r(s,t)-r)\nonumber \\
 & \times & \int\mathcal{D\zeta}(s,t)\exp\Bigg(-\frac{1}{2}\sum\int_{0}^{N}ds\zeta^{2}(s,t)+i\sum_{t}\int_{0}^{N}dsq(s,t)\Bigg(r(s,t+\varepsilon)-r(s,t)\nonumber \\
 & - & D\beta\varepsilon\left(\frac{3}{a^{2}}\frac{d^{2}r}{ds^{2}}-\int dr'\nabla_{r}v(r(s,t)-r')\rho(r',t)\right)-\sqrt{2D\varepsilon}\zeta(s,t)\Bigg)\Bigg)
\end{eqnarray}
The term $q(s,t)$ in the functional integral can be written as a functional derivative w.r.t. the noise field $\zeta$ as
\begin{eqnarray}
\frac{\delta Q}{\delta\rho(r,t)} & = & -\beta\sqrt{D\varepsilon/2}\int\mathcal{D}r(s,t) \mathcal{D}q(s,t)e^{-i\varepsilon\sum_{t}\int ds\phi(r(s,t),t)}\int_{0}^{N}ds\nabla v(r(s,t)-r)\times\nonumber \\
 &  & \int\mathcal{D\zeta}(s,t)\exp\Bigg(-\frac{1}{2}\sum\int_{0}^{N}ds\zeta^{2}(s,t)\Bigg) \nonumber \\
& & \frac{\delta}{\delta\zeta(s,t)}\exp\Bigg(i\sum_{t}\int_{0}^{N}ds \: q(s,t)\Bigg(r(s,t+\varepsilon)-r(s,t)\nonumber \\
 &  &  -D\beta\varepsilon\left(\frac{3}{a^{2}}\frac{d^{2}r}{ds^{2}}-\int dr'\nabla_{r}v(r(s,t)-r')\rho(r',t)\right)-\sqrt{2D\varepsilon}\zeta(s,t)\Bigg)\Bigg)
\end{eqnarray}
Integrating (functionally) by parts the derivative with respect
to the noise $\zeta(s,t)$ we obtain
\begin{eqnarray}
\frac{\delta Q}{\delta\rho(r,t)} & = & -\beta\sqrt{D\varepsilon/2}\int\mathcal{D}r(s,t) \mathcal{D}q(s,t)e^{-i\varepsilon\sum_{t}\int ds\phi(r(s,t),t)}\int_{0}^{N}ds\nabla v(r(s,t)-r)\times\nonumber \\
 &  & \int\mathcal{D\zeta}(s,t)\zeta(s,t)\exp\Bigg(-\frac{1}{2}\sum\int_{0}^{N}ds\zeta^{2}(s,t)\Bigg)\exp\Bigg(i\sum_{t}\int_{0}^{N}ds \: q(s,t)\Bigg(r(s,t+\varepsilon)\nonumber \\
 &  & -r(s,t) -D\beta\varepsilon\left(\frac{3}{a^{2}}\frac{d^{2}r}{ds^{2}}-\int dr'\nabla_{r}v(r(s,t)-r')\rho(r',t)\right)-\sqrt{2D\varepsilon}\zeta(s,t)\Bigg)\Bigg)
\end{eqnarray}
which can be finally rewritten in the simple form
\begin{eqnarray}
\frac{\delta Q}{\delta\rho(r,t)} & = & -\beta\sqrt{D\varepsilon/2}\int_{0}^{N}ds \; <\zeta(s,t)\nabla v(r(s,t)-r)>_{Q}\label{expectation}
\end{eqnarray}

Because of the Ito discretization used for the Langevin equation,
$r(s,t)$ depends on the noise variable $\zeta(s,t-\varepsilon)$ at times earlier or equal to $t-\varepsilon$ and not on $\zeta(s,t)$. Therefore, we have {the decoupling}\color{black}
\begin{eqnarray}
\label{decoupling}
<\zeta(s,t)\nabla v(r(s,t)-r)>_{Q}&=&<\zeta(s,t)><\nabla v(r(s,t)-r)>_{Q} \nonumber\\
&=& 0
\end{eqnarray}
since the expectation value of $\zeta(s,t)$ is equal
to 0.

At this mean-field level, the second mean-field equation (\ref{phi}) thus becomes
\[
\phi(r,t)=0.
\]
Therefore, in the mean-field approximation, the dynamics of the polymer
system can be described in terms of a single-chain dynamics:
\begin{equation}
r(s,t+\varepsilon)=r(s,t)+D\beta\varepsilon\left(\frac{3}{a^{2}}\frac{d^{2}r}{ds^{2}}-\int dr'\nabla_{r}v(r(s,t)-r')\rho(r',t)\right)+\sqrt{2D\varepsilon}\zeta(s,t)\label{MF1}
\end{equation}
where the polymer density field is given by
\begin{equation}
\rho(r,t)=M\int_{0}^{N}ds<\delta(r(s,t)-r)>_{Q},\label{MF2}
\end{equation}
the right hand side being computed as an average over the same independent, single chain dynamics generated by the functional $Q$:
\begin{eqnarray}
Q(\rho,\phi) & = & \int\mathcal{D}r(s,t)<\prod_{t}\delta\Bigg(r(s,t+\varepsilon)-r(s,t) \nonumber \\
&-&D\beta\varepsilon\left(\frac{3}{a^{2}}\frac{d^{2}r}{ds^{2}}-\int dr'\nabla_{r}v(r(s,t)-r')\rho(r',t)\right)-\sqrt{2D\varepsilon}\zeta(s,t)\Bigg)>\label{MF3}
\end{eqnarray}

A few remarks can be made at this stage. First of all, eqs.(\ref{MF1}) and (\ref{MF2}) represent a very significant simplification of the original microscopic model with $M$ chains. In particular the mean-field dynamics constitutes a stochastic dynamics for a \emph{single chain} whose monomers experience a force $- \nabla_r w (r ,t)$ produced by a mean-field $w(r,t)$ given by
$$
w(r,t) = \int dr^\prime \: v(r - r^\prime ) \rho (r^\prime ,t).
$$
The field is in turn determined instantaneously by the average density $\rho (r,t)$ computed from eq.(\ref{MF2}). This is an entirely intuitive result that could be obtained from the starting many-chain dynamics by approximating the exact fluctuating molecular field
$$
\hat{w}(r,t) = \int dr^\prime \: v(r - r^\prime ) \hat\rho (r^\prime ,t)
$$
by the average field
$$
w(r,t) = < \hat{w} (r,t)> = \int dr^\prime \: v(r - r^\prime ) < \hat\rho (r^\prime ,t) >
$$
and then recognizing that since the chains are now evolving independently of one another, $< \hat\rho ( r,t) >$ can be replaced by $\rho (r,t)$ given by eq.(\ref{MF2}).

A second important point is that the average density at time $t$ in eq.(\ref{MF2}) depends on the statistical properties of the chain configuration $r(s,t)$ at time $t$, which according to the mean-field Langevin eq.(\ref{MF1}), depends on $\rho(r,t-\varepsilon)$ and not $\rho(r,t)$. Thus, there is no self-consistency required in computing the density $\rho(r,t)$. The density can be simply evaluated after each time step in the single-chain Langevin dynamics, albeit by propagating enough replicas of the chain to accurately evaluate the $Q$-average on the r.h.s. of eq.(\ref{MF2}) (see discussion below).

Finally, we point out that our derivation could be easily generalized to multiple polymer species, to other architectures, e.g. block copolymers, and to semi-flexible chains. In the latter case, within a Gaussian chain model we could simply add a bending energy
term such as $\frac{\kappa}{2}\int_{0}^{N}ds\left(\frac{d^{2}r}{ds^{2}}\right)^{2}$ to the Hamiltonian, which would result in an extra force
term $-\kappa\frac{d^{4}r}{ds^{4}}$ in
the single-chain Langevin equation above. If, in addition, the interactions depend on the tangent vector $u_{k}(s,t)$ of the chain segment
at $r_{k}(s,t)$, then the mean-field density will be a higher dimensional object $\rho (r,u,t)$ depending on both position and orientation.

{ An important last comment is in order at this stage: had we used the Stratonovich discretization of the Langevin equation
\begin{eqnarray}
r_{k}(s,t+\varepsilon)&=&r_{k}(s,t)+\frac{D}{2}\beta\varepsilon \Bigg ( \frac{3}{a^{2}}\frac{d^{2}r_{k}(s,t)}{ds^{2}}-\sum_{l}\int_{0}^{N}ds'\nabla_{k}v(r_{k}(s,t)-r_{l}(s',t)) \nonumber \\
&+& \frac{3}{a^{2}}\frac{d^{2}r_{k}(s,t+\varepsilon)}{ds^{2}}-\sum_{l}\int_{0}^{N}ds'\nabla_{k}v(r_{k}(s,t+\varepsilon)
-r_{l}(s',t+\varepsilon))\Bigg)
\nonumber \\
&+&\sqrt{2D\varepsilon}\zeta_{k}(s,t)
\end{eqnarray}
which involves implicitly the position of the chain at time $t+\varepsilon$ and at time $t$,
the definition of the probability distribution eq.(\ref{proba}) would have entailed a Jacobian, but more importantly, the implicit character of this equation would have prevented the decoupling of eq.(\ref{decoupling}) that renders the mean-field equations so simple and intuitive. In fact, although  Ito and Stratonovich discretizations have the same continuous limit when treated exactly, such agreement is not evident at the mean-field level, with the Stratonovich form leading to intractable equations.
}
\color{black}
\section{Convergence to the SCFT}

We next consider the limit of large time. Assume that the density converges
to an equilibrium density $\lim_{t\rightarrow\infty}\rho(r,t)=\rho_{0}(r)$.
To simplify the notation, we return to the continuous notation in time.
At large time, the mean-field Langevin equation becomes
\begin{equation}
\frac{dr(s,t)}{dt}=D\beta\left(\frac{3}{a^{2}}\frac{d^{2}r}{ds^{2}}-\int dr'\nabla_{r}v(r(s,t)-r')\rho_{0}(r')\right)+\eta(s,t)\label{stat1}
\end{equation}
where $\rho_{0}(r)$ is the average density generated by the different
realizations of eq. (\ref{stat1}). It is well-known that in the long-time limit, the probability distribution of
the variable $r(s,t)$ generated by (\ref{stat1}) is the Boltzmann
distribution associated with the Hamiltonian of that equation. The
Hamiltonian is given by
\[
U=\frac{3}{2a^{2}}\int_{0}^{N}ds\left(\frac{dr}{ds}\right)^{2}+\int dr'\int_{0}^{N}ds\, v(r(s)-r')\rho_{0}(r')
\]
Equation (\ref{MF2}) for $\rho_{0}$ thus reduces to
\[
\rho_{0}(r)=M \frac{\int\mathcal{D}r(s)\int_0^N ds \delta(r-r(s))e^{-\beta U}}{\int\mathcal{D}r(s)e^{-\beta U}}
\]
or
\[
\rho_{0}(r)=\frac{M}{Z}\int_{0}^{N}ds\int\mathcal{D}r(s)\delta(r-r(s))e^{-\frac{3}{2a^{2}}\int_{0}^{N}ds\left(\frac{dr}{ds}\right)^{2}-\int_{0}^{N}ds\Phi(r(s))}
\]
where $Z$ is the single chain partition function and $\Phi (r)$ is a static mean-field
\[
\Phi(r)=\int dr'v(r-r')\rho_{0}(r').
\]

Using standard quantum mechanical notations, we define the Hamiltonian
\[
H=-\frac{a^{2}}{6}\nabla^{2}+\Phi(r)
\]
It follows that the density can be written as
\[
\rho_{0}(r)=M\int_{0}^{N}ds\frac{\int\int dr_{1}dr_{2}<r_{1}|e^{-(N-s)H}|r><r|e^{-sH}|r_{2>}}{\int\int dr_{1}dr_{2}<r_{1}|e^{-NH}|r_{2>}}
\]
Defining the usual ``chain propagator'' fields $\phi$ and $\phi^{*}$ by
\[
\phi(r,s)=\frac{1}{\sqrt{Z}}\int dr_{2}<r|e^{-sH}|r_{2}>
\]
and
\[
\phi^{*}(r,s)=\frac{1}{\sqrt{Z}}\int dr_{1}<r_{1}|e^{-(N-s)H}|r>
\]
we havev
\[
\rho_{0}(r)=M\int_{0}^{N}ds \: \phi^{*}(r,s)\phi(r,s)
\]
The propagator fields $\phi$ and $\phi^{*}$ satisfy the diffusion equations
\[
\left(\frac{\partial}{\partial s}+H\right)\phi=0
\]
and
\[
\left(-\frac{\partial}{\partial s}+H\right)\phi^{*}=0
\]
These last equations are just the usual SCFT equations\cite{ghfbook}. Therefore, if our dynamical mean-field equations converge to an equilibrium distribution of segments, we have proved that the distribution is that implied by the static mean-field theory -- SCFT.

\section{Discussion and numerical implementation}
\label{discussion}
The practical implementation of the mean-field equations (\ref{MF1}) and (\ref{MF2}) is very straightforward. In order to compute the ensemble average in eq. (\ref{MF2}), we have to choose the number of chain replicas that will be used for the sampling. Let us denote this number by $M_S$ to distinguish it from the physical number of chains in the starting model $M$. The possibility that accurate sampling could be done with $M_S \ll M$ offers potential advantage to the dynamical mean-field theory over a full $M$-chain simulation.

In our discussion of numerical implementation, we only make explicit the time discretization. Space variables as well as the curvilinear coordinate $s$ must also be discretized, but as this is already standard in SCFT and SCMF and can be done many different ways, for the sake of simplicity we will retain the continuous notation for $\{r\}$ and $\{s\}$. Each of the $M_S$  replica chains is evolved according to the same eq. (\ref{MF1}), but with a different noise history $\zeta(s,t)$. To be more specific, assume we have generated the $M_S$ chain samples up to time $t$ according to eq. (\ref{MF1}). We thus know the configurations $\{r_{\alpha}(s,t)\}$ for all the chains of the sample ensemble $\alpha \in \{1,...,M_S\}$ and any monomer $s$ up to time $t$. Equation (\ref{MF2}) can then be applied in the form
\begin{equation}
\rho(r,t)=\frac{M}{M_S} \int_{0}^{N}ds\sum_{\alpha=1}^{M_S} \delta(r_{\alpha}(s,t)-r) \label{MF6}
\end{equation}
where the local error in this expression is expected to be of order $(V/[M_S N \Delta V])^{1/2}$, where $V$ is the volume and $\Delta V$ is the cell volume used for the spatial discretization.
We can then use this calculated $\rho(r,t)$ and the current chain replica configurations $\{r_{\alpha}(s,t)\}$ in eq. (\ref{MF1}) to compute the configuration of all replicas at the next time step, $\{r_{\alpha}(s,t+\varepsilon)\}$. Of course, the whole procedure is initiated by generating an initial set of $M_S$ replica chains $\{r_{\alpha}(s,0)\}$ at time $t=0$.

Clearly, the time evolution of the $M_S$ replicas over a single time step can be trivially parallelized since all chains are independent from each other. However, they are coupled at each time step through the density field $\rho(r,t)$ according to eq. (\ref{MF6}). The procedure of updating the density breaks the parallelization, although the contributions to the density from each chain replica can be computed on separate processors before being gathered and summed, the latter merge steps being relatively inexpensive compared with replica propagation.  Anticipating that the density evolves more slowly than the chain coordinates, it is further desirable that the density updates be performed only every $n$ time steps, where $n \ge 1$ is an integer determined by the targeted numerical accuracy of the dynamical trajectory.

At this stage it is worth pointing out that the proposed dynamical mean-field algorithm with the particular choice of $M_S = M$ corresponds exactly to the SCMF procedure of M\"{u}ller and coworkers\cite{mueller1,mueller2}, with the slight modification that M\"{u}ller et. al. substitute kinetic MC moves for the single-chain Langevin dynamics. We further note that the SCMF case of $M_S = M$ and $n=1$ (density updates every time step) corresponds exactly to the full $M$-chain Langevin dynamics of eq.({\ref{density_langevin}), has the same computational complexity, and is evidently \emph{not a mean-field theory} as the computed densities and non-bonded forces fluctuate in accordance with the local molecular environment. Daoulas and M\"{u}ller\cite{mueller2} argue that for practically useful values of $n > 1$, the SCMF procedure yields realistic field fluctuations at equilibrium, numerically validated by a comparison with full MC simulations. Out of equilibrium, the error incurred in the dynamical trajectories by choosing $n>1$ is difficult to assess \textit{a priori}, and must be validated in specific situations by comparing with full many-chain simulations.

Returning to the dynamical mean-field theory, a natural question to address is how large should $M_S$ be compared to $M$. Of course, the smaller $M_S$, the faster the algorithm. However, $M_S \ll M$ could produce unphysically large fluctuations in $\rho$ and a significant departure from the mean-field dynamical trajectory due to large sampling error of the r.h.s. of eq (\ref{MF2}).  One might hope that parameters $M_S < M$ and $n>1$ could be identified, e.g. in dense systems of long polymers, whereby efficient simulations could be conducted of sufficient accuracy. To the extent that $M_S$ is less than $M$, there would be a proportional reduction in computational effort compared with the strict SCMF algorithm. We look forward to numerical investigations that explore this issue. Finally, we emphasize that the (forward Euler) Langevin single-chain dynamics used for the present analysis could be readily replaced by a ``smart'' or force-biased kinetic MC scheme, undoubtedly allowing for larger time steps and better performance.

In summary, we have derived a dynamical mean-field theory for polymeric fluids based on a saddle point approximation to a functional integral description of many-chain dynamics. The theory reduces the full many-chain dynamics to a much simpler problem involving the coupled stochastic dynamics of a \emph{single chain} in a time-dependent, ensemble averaged, mean-field determined by the average density $\rho (r,t)$. Remarkably, the relationship between the mean-field and the density is instantaneous in time, as assumed in the recently developed single chain in mean-field (SCMF) approaches, and is non-local in space only to within the range of the potential. Furthermore, at least for the microscopic model considered, the analysis does not rely on the identification of additional slow collective variables beyond the density, such as conformational stress, but embeds such dynamical information through the retained single chain degrees of freedom. The most natural algorithm for implementing the resulting theory is to independently propagate $M_S$ replicas of the single chain, e.g. by a Langevin or MC procedure, and periodically update the mean-field density $\rho$ appearing in the single chain equations. In the special case of $M_S$ equal to the number of chains $M$ in the corresponding many-chain model, our procedure reduces exactly to the SCMF approach of Muller and coworkers.

\begin{acknowledgments}
This work was supported by the Condensed Matter and Materials Theory Program of the NSF under award No.\ DMR 1160895. We are grateful to M. M\"{u}ller for helpful discussions.
\end{acknowledgments}

\end{document}